\documentclass[tighten,twocolumn]{aastex62}
\accepted{10/10/18}
\received{9/10/18}
\submitjournal{AJ}
\usepackage{graphicx}
\usepackage{natbib}
\usepackage{latexsym}
\usepackage{amssymb}
\usepackage{longtable}
\usepackage{amsmath}
\usepackage{url}
\def\tp{$T(\text{P})$}

\def\msun{\ifmmode {\rm\,M_\odot}\else ${\rm\,M_\odot}$\fi}
\def\Msun{\ifmmode {\rm\,\it{M_\odot}}\else ${\rm\,M_\odot}$\fi}
\def\lsun{\ifmmode {\rm\,L_\odot}\else ${\rm\,L_\odot}$\fi}
\def\Lsun{\ifmmode {\rm\,\it{L_\odot}}\else ${\rm\,L_\odot}$\fi}
\def\rsun{\ifmmode {\rm\,R_\odot}\else ${\rm\,R_\odot}$\fi}
\def\Rsun{\ifmmode {\rm\,\it{R_\odot}}\else ${\rm\,R_\odot}$\fi}
\def\Tsun{\ifmmode {\rm\,T_\odot}\else ${\rm\,T_\odot}$\fi}
\def\arcsec{\ifmmode {^{\prime\prime}}\else $^{\prime\prime}$\fi}
\def\asec{\ifmmode {^{\prime\prime}}\else $^{\prime\prime}$\fi}
\def\arcmin{\ifmmode {^{\prime}}\else $^{\prime}$\fi}
\def\amin{\ifmmode {^{\prime}}\else $^{\prime}$\fi}
\def\simlt{\mathrel{\spose{\lower 3pt\hbox{$\mathchar"218$}}
     \raise 2.0pt\hbox{$\mathchar"13C$}}}
\def\simgt{\mathrel{\spose{\lower 3pt\hbox{$\mathchar"218$}}
\     \raise 2.0pt\hbox{$\mathchar"13E$}}}

\begin{document}

\title{Evidence of magnetic star-planet interactions in the HD 189733 system
from orbitally-phased \ion{Ca}{2} K variations}

\author{P. Wilson Cauley}
\affiliation{Arizona State University, School of Earth and Space Exploration, Tempe, AZ 85287}

\author{Evgenya L. Shkolnik}
\affiliation{Arizona State University, School of Earth and Space Exploration, Tempe, AZ 85287}

\author{Joe Llama}
\affiliation{Lowell Observatory, Flagstaff, AZ 86001}

\author{Vincent Bourrier}
\affiliation{Observatoire de l'Universit\'{e} de Gen\`{e}ve, 1290 Sauverny, Switzerland}

\author{Claire Moutou}
\affiliation{CFHT Corporation, Kamuela, HA 96743}

\correspondingauthor{P. Wilson Cauley}
\email{pwcauley@gmail.com}

\begin{abstract} 

Magnetic star-planet interactions (SPI) provide a detection method and insight
into exoplanet magnetic fields and, in turn, exoplanet interiors and
atmospheric environments.  These signatures can be sporadic and difficult to
confirm for single-epoch observations of a system due to inhomogeneous stellar
magnetospheres and periodic variability in stellar magnetism. Thus an ideal SPI
search consists of multiple epochs containing observations on consecutive
nights spanning at least one complete planetary orbit. Such data sets are rare
but do exist for some of the most intensely studied hot Jupiter systems. One
such system is HD 189733 for which six suitable SPI data sets exist, the result
of spectroscopic monitoring to perform some of the first SPI searches and also
to study the star's magnetic field. Here we perform a uniform analysis of six
archival \ion{Ca}{2} K data sets for HD 189733, spanning 2006 June through 2015
July, in order to search for magnetic SPI signatures in the chromospheric line
variations. We find significant evidence for modulations of \ion{Ca}{2} K with
a $2.29\pm0.04$ day period in the 2013 August data, which is consistent with
the planet's orbital period. The peak in the orbital variations occurs at
$\phi_\text{orb} \approx 0.9$, which corresponds to the SPI emission leading
the planet with a phase difference of $\Delta\phi \approx 40^\circ$ from the
sub-planetary point. This is consistent with the phase lead predictions of
non-linear force-free magnetic field SPI models.  The stellar magnetic field
strength at the planet's orbit is greatest in 2013 August which, due to the
energy released in magnetic SPI scaling with $B_*$, lends strength to the SPI
interpretation. 

\end{abstract}

\keywords{}

\section{Introduction}
\label{sec:intro}

Close-in massive planets, i.e., those with $M_\text{p} \gtrsim 0.5 M_\text{J}$
and $a_\text{orb} \lesssim 10 R_*$, are capable of strongly interacting with
their host stars via tides or reconnection processes involving the
magnetospheres of both objects \citep[e.g.,][]{cuntz00,cranmer07,lanza09,shkolnik17},
potentially enhancing the activity level of the host star and affecting the
spin evolution of the star and orbital evolution of the planet
\citep{lanza10,cohen11,strugarek15}. Magnetic star-planet interactions, or SPI,
can reveal information about exoplanetary atmospheres and interiors through the
detection of magnetic fields \citep[e.g.,][]{lanza09,lazio16}, characteristics
which are difficult to probe and about which we currently have limited
knowledge. 

HD 189733 b is one of the most well-studied hot Jupiters due to its host star's
brightness, high activity level, and a favorable $(R_\text{p}/R_*)^2$ for
atmospheric characterization \citep{bouchy05}. A variety of experiments have
been performed searching for SPI signatures in the system. \citet{shkolnik08}
presented evidence for variability in the \ion{Ca}{2} H and K lines as a
function of the planet's orbital period, a typical manifestation of magnetic
SPI \citep{shkolnik03,shkolnik05}. Similar variations were searched for in
\citet{fares10} across two epochs of $\approx 20$ nights each. They found no
clear evidence for a SPI signal at the planet's orbital period.
\citet{pillitteri11}, \citet{pillitteri14}, and \citet{pillitteri15} presented
evidence of X-ray and UV flares associated with a narrow range of planetary
orbital phase, evidence of magnetic SPI between the stellar corona and
planetary magnetic field or, in the case of the UV flares, accreting planetary
material onto the stellar surface.  \citet{poppenhaeger14} looked at X-ray
emission from both HD 189733 A, the planet host, and HD 189733 B, a distant M4V
companion, and found that HD 189733 A's X-ray flux suggested a younger age than
its companion. They surmised that the enhanced X-ray flux and fast rotation is
a result of stellar spin-up by HD 189733 b.  Finally, \citet{cauley17}
demonstrated that the H$\alpha$ line core flux surrounding the planetary
transit showed abnormal variations compared with times far from transit,
suggestive of either SPI or absorption by circumplanetary material.

Magnetic SPI signals can be sporadic: changes in the relative orientation of
the stellar and planetary magnetic fields, as well as longer term variability
in the stellar magnetic field strength and topology, can cause the SPI
mechanism to turn on and off over orbital and stellar activity cycles
\citep{lanza10,cohen11,strugarek15}. Furthermore, stochastic changes in stellar
activity, which are common for active stars such as HD 189733, unrelated to the
planet can contaminate night-to-night signatures, potentially masking phased
SPI emission or even resulting in such a signal. Stochastic variability can be
tested as the culprit of orbitally phased emission by producing random light
curves at the observed phases and calculating the probability of measuring the
real signal. This highlights the need for observations which span multiple
planetary orbits and, ideally, stellar activity cycles and statistically
testing any orbitally phased variability.  

In the past decade, numerous observing campaigns have been executed with the
purpose of mapping the magnetic field strength and topology of HD 189733 A
\citep{fares10,fares17}, as well as a four night SPI investigation by
\citet{shkolnik08}. These data are ideally formatted to search for SPI signals:
high signal-to-noise \ion{Ca}{2} H and K spectra obtained regularly across
multiple planetary orbits.  In this paper we present a uniform search for SPI
signals in these archival \ion{Ca}{2} K spectra of HD 189733. The data consist
of six observing runs each containing at least 8 nights, a duration of time
which spans $\geq 3$ orbital periods of HD 189733 b ($P_\text{orb} = 2.2$
days). Although previously analyzed for SPI, we include the data sets from
\citet{fares10} for completeness and to apply a consistent analysis.
\autoref{sec:obs} details the observations and summarizes the data reduction,
including steps taken to produce the residual flux spectra used in the SPI
analysis. The results are laid out in \autoref{sec:results} and discussed
in \autoref{sec:discussion}. A brief summary
of our conclusions are presented in \autoref{sec:conclusion}.

\section{Observations and data reduction}
\label{sec:obs}

We have re-purposed archived high resolution high-quality spectroscopy of the
HD 189733 hot Jupiter system. The selected epochs are detailed in
\autoref{tab:obs}.  All of the data were taken from the PolarBase archive
\citep{petit14}. The spectra were reduced using the automated pipeline
\texttt{libre-esprit}, fully described in \citet{donati97} and to which we
refer the reader for details.  Briefly, each exposure is bias subtracted and
flat fielded to remove pixel-to-pixel variations. Inter-order scattered light
is then approximated and removed. The spectra are optimally extracted and a
wavelength solution is computed using Th-Ar comparison exposures.  The
normalized pipeline-reduced spectra are then corrected for heliocentric
velocities and small wavelength shifts between spectra are adjusted via a
$\chi^2$ comparison of narrow spectral lines in the \ion{Ca}{2} H and K order. 

\begin{deluxetable*}{cccccccc}
\tablewidth{1.99\textwidth}
\tablecaption{Observations of HD 189733\label{tab:obs}}
\tablehead{\colhead{Observation epoch}&\colhead{UT start}&\colhead{UT end}&\colhead{$N_\text{nights}$}&\colhead{Instrument}&
\colhead{Telescope}&\colhead{$\lambda/\Delta\lambda$}&\colhead{References}}
\colnumbers
\tabletypesize{\scriptsize}
\startdata
 2006 Jun$^\text{a}$ & 2006 Jun 10 & 2006 Jun 13 & 4 & ESPaDOns & CFHT & 65,000 & \citet{moutou07}\\
 	        & 2006 Jun 16 & 2006 Jun 19 & 4 & ESPaDOns & CFHT & 80,000 & \citet{shkolnik08} \\
 2007 Jun$^\text{b}$ & 2007 Jun 23 & 2007 Jul 4 & 8 & ESPaDOns & CFHT & 65,000  & \citet{fares10} \\
 2008 Jul & 2008 Jul 15 & 2008 Jul 24 & 8 & NARVAL & TBL & 65,000  & \citet{fares13} \\
 2013 Aug & 2013 Aug 4 & 2013 Aug 22 & 11 & NARVAL & TBL & 65,000 & \citet{fares17} \\
 2013 Sep & 2013 Sep 2 & 2013 Sep 24 & 13 & NARVAL & TBL & 65,000 & \citet{fares17} \\
 2015 Jul & 2015 Jul 6 & 2015 Jul 16 & 10 & NARVAL & TBL & 65,000 & \citet{fares17} \\
\enddata
\tablenotetext{\text{a}}{The 2006 June epoch is a combination of two 4-night observing runs. The $R\approx80,000$
data is convolved with a Gaussian kernel down to $R\approx65,000$.}
\tablenotetext{\text{b}}{The average spectrum from 2007 June 30 is discarded due to low S/N.}
\end{deluxetable*}

The optimal strategy for detecting SPI signatures as a function of planetary
orbital phase is to repeatedly observe the host star on consecutive nights
across at least two full orbital periods. For this reason, we require that an
epoch of observations contains $\geq 5$ exposures taken across at least two
full planetary orbits. Sparsely sampled portions of observing runs are excluded
in favor of stretches with continuous data or that only contain gaps less than
or similar to a single orbital period. This can help mitigate contamination
from changing stellar activity levels that are not related to rotational
modulation.  For example, we exclude the first four nights of the full 2007
June data set and the first night of the 2008 July data set due to sparse
sampling and their separation from the more densely sampled portions. All of
the epochs consist of consecutive observations, with the one exception for 2007
June noted in \autoref{tab:obs}. 

Individual nightly spectra are examined for quality and rejected if the
standard deviation of the residual intra-night flux of the spectrum, measured
outside of the \ion{Ca}{2} K line core, is $>1.5\times$ the mean residual flux
standard deviation for that night. We do not include an analysis of \ion{Ca}{2}
H since it correlates directly with \ion{Ca}{2} K and is the weaker member of
the doublet.

\subsection{Preparation of the spectra for SPI analysis}
\label{sec:spiprep}

{Here we explicitly lay out the steps taken to produce the
residual \ion{Ca}{2} K fluxes, which is the data product used in the
SPI analysis. The individual steps taken, which are illustrated in
\autoref{fig:procsteps}, are as follows:

\begin{enumerate}

\item Precise normalization of the \ion{Ca}{2} K line core was done using a
linear fit to the 0.2 \AA\ on either end of a 6 \AA\ range centered on 3933.66
\AA

\item The median was taken of the normalized spectra for each night

\item The mean of the nightly median spectra was taken to create an
average \ion{Ca}{2} K spectrum for the epoch

\item The mean spectrum was subtracted from each night's median to create
residual \ion{Ca}{2} K profiles for each night

\item Small offsets in the residual profiles were corrected using a low-order
spline fit to points outside of the central 1 \AA\ in the line core 

\item The flux were summed across the central 1 \AA\ to produce the
residual \ion{Ca}{2} K flux for that night

\item A sinusoidal function was fit to the residual fluxes to remove variations
due to stellar rotation

\end{enumerate}

\begin{figure*}[htbp]
   \centering
   \includegraphics[scale=.76,clip,trim=9mm 10mm 10mm 15mm,angle=0]{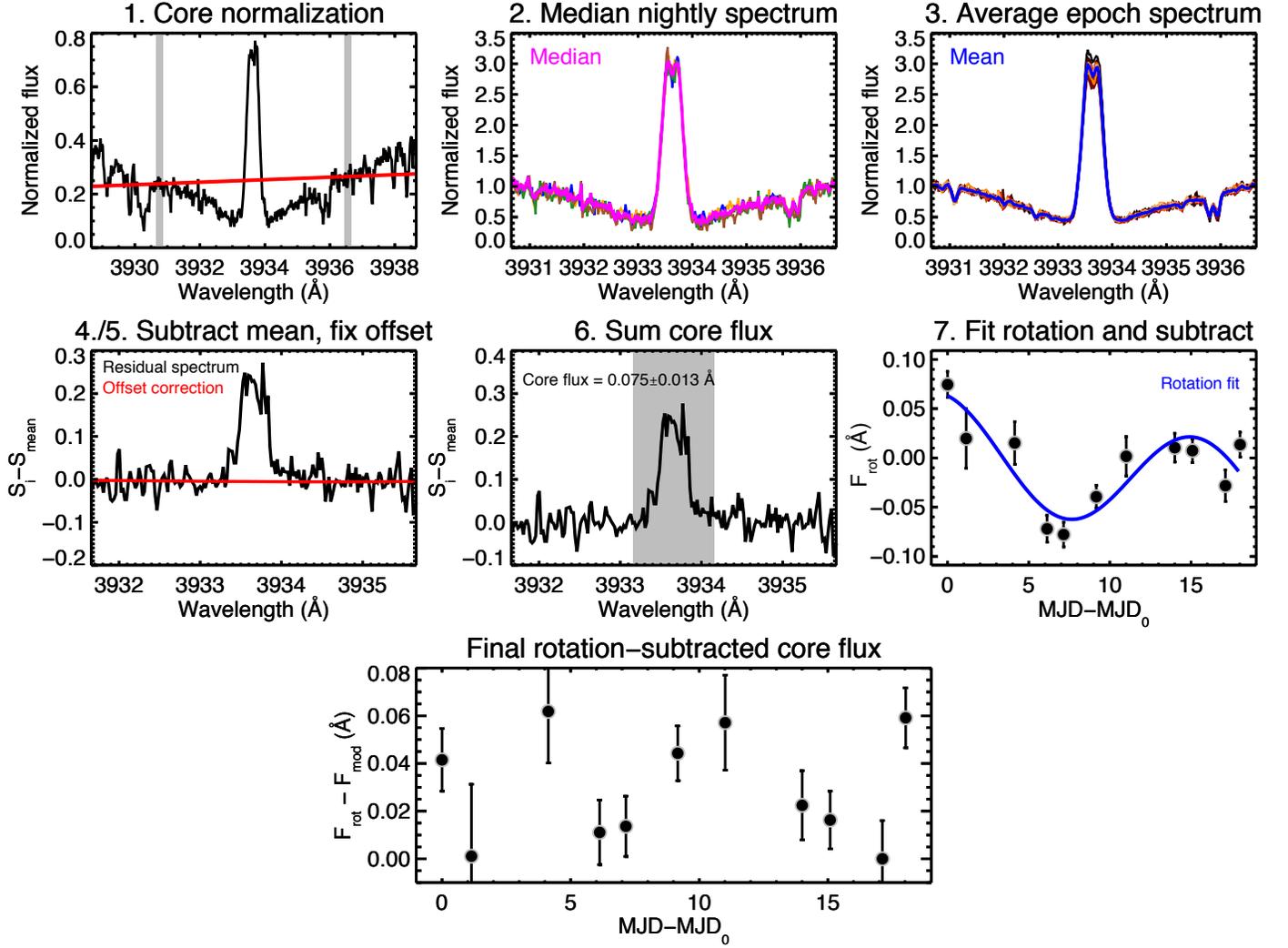}

   \figcaption{Illustration of the step-by-step procedure for deriving the residual
\ion{Ca}{2} K flux. The data from 2013 August are used as the examples. \label{fig:procsteps}}

\end{figure*}

Each individual spectrum was normalized using a 6 \AA\ region centered on the \ion{Ca}{2} K
line. A line was then fit to the average of the points
comprising 0.2 \AA\ on either end of the 6 \AA\ range (Step 1 in \autoref{fig:procsteps}). 
The median of the normalized spectra for each night was taken to produce the median nightly 
spectrum (Step 2 in \autoref{fig:procsteps}). 
The median spectrum was used instead of the mean since intra-night variability 
is due to stochastic stellar activity, a non-Gaussian process. All median nightly
spectra from the epoch were then averaged to create a mean \ion{Ca}{2} K
profile (Step 3 in \autoref{fig:procsteps}). Mean spectra are shown in the top row 
of \autoref{fig:profs} for each epoch. 

Residual profiles for each night were created by subtracting the mean epoch spectrum
from the median nightly spectra (Step 4 in \autoref{fig:procsteps}). Any additional slope or
offset in the residual spectrum continuum was subtracted with a low-order spline to
points outside of the line core. These offsets
were typically $\approx 10\times$ smaller than the peaks of the residual lines
cores (e.g., the red line in the middle-left panel of \autoref{fig:procsteps}). The
residual core flux was summed across a 1 \AA\ region centered on the \ion{Ca}{2} K
rest wavelength (Step 6 in \autoref{fig:procsteps}). The uncertainty for each
flux value in the residual spectrum was set equal to the standard deviation of
the spectrum outside of the line core (gray region in the middle panel of 
\autoref{fig:procsteps}). The final step was removing variations due to stellar rotation,
which is outlined in detail in \autoref{sec:rotvar}.

\subsection{Removing stellar rotational variations}
\label{sec:rotvar}

Before assessing the presence of SPI variations, rotational modulation on the
star, if present, must be approximated and removed.  HD 189733 is an active
star and shows clear rotational variability \citep[e.g.,][]{hebrard06,fares10}.
There is also evidence that HD 189733 is rotating differentially: using the
2007 June data set included here, \citet{fares10} measured a differential
rotation rate of $\text{d}\Omega = 0.146\pm0.049$ radians day$^{-1}$, which
corresponds to an equatorial rotation period of $11.94\pm0.16$ days and a polar
rotational period of $16.53\pm2.43$ days. These measurements were confirmed by
\citet{fares17} who found $\text{d}\Omega = 0.11\pm0.05$ radians day$^{-1}$ and
an equatorial rotational period of $11.7\pm0.1$ days. \citet{cegla16} also
place a lower limit on the differential rotation value of $\text{d}\Omega \geq
0.12$ radians day$^{-1}$.

We model the rotational variations in the residual \ion{Ca}{2} K flux as a
sinusoid with a linear trend, letting the period of rotation vary as a free
parameter \citep[e.g.,][]{fares10}. We note that models with $2-3$ sinusoids, which simulate active
regions across multiple differentially rotating latitudes, do not significantly
improve the rotational fits. The linear trend is included to approximate active
regions of varying strength rotating into view or changes in emission strength
during a rotation.  A single sinusoid cannot account for such changes. The
rotational variability function has the form

\begin{equation}\label{eq:strot}
F = A_0 \sin\big((t-T_0) 2 \pi / P_\mathrm{rot}\big) + bt + C
\end{equation}

\noindent where $A_\text{0}$ is the amplitude, $T_\text{0}$ is the time offset
from the date of the first observation, $P_\text{rot}$ is the rotation period,
and $bt+C$ is a linear trend.  The fits are performed using a Markov
chain Monte Carlo procedure based on the algorithm from \citet{goodman10}
\citep[see also][]{foreman12}. With the exception of the rotation period,
non-restrictive uniform priors are set for all of the parameters. The values
allowed for the rotation period are uniformly restricted to $11.0 \leq
P_\text{rot} \leq 16.5$ days according to the measurements from
\citet{fares17}. This allows us to account for rotating active regions at
latitudes other than the equator, effectively simulating differential
rotation but with the constraint that only one latitude is producing the
variations. We take the median of the marginalized
posterior distributions as the best-fit value for each parameter. The
best-fitting rotational curve is then subtracted from the raw residual flux
curve, which allows the resulting variations to be examined for potential SPI
signatures (e.g., bottom panel in \autoref{fig:procsteps}). 

The rotational parameters and their $68\%$ confidence intervals are given in
\autoref{tab:rotfits}. One interesting result to note about the best-fit periods
is their non-equatorial values. The rotation periods derived here, combined
with the differential rotation measurements of \citet{fares10,fares17}, suggest
active regions located at latitudes of $\approx 40^\circ-80^\circ$. This is
consistent with the active region latitudes found by \citet{lanza11} while
modeling spots and activity-induced radial velocity variations. 

\begin{deluxetable}{cccccc}
\tablecaption{Stellar rotation fit parameters from \autoref{eq:strot}\label{tab:rotfits}}
\tablehead{\colhead{}&\colhead{$P_\text{rot}$}&\colhead{}&
\colhead{$T_0$}&\colhead{}&\colhead{}\\
\colhead{Epoch}&\colhead{(days)}&\colhead{$A_0$}&
\colhead{(days)}&\colhead{$C$}&\colhead{$b$}}
\colnumbers
\tabletypesize{\scriptsize}
\startdata
2006 Jun  & $14.9^{+1.1}_{-2.5}$   & $0.055^{+0.018}_{-0.020}$   & $1.43^{+0.96}_{-0.82}$   & $-0.057^{+0.027}_{-0.027}$   & $0.010^{+0.005}_{-0.005}$  \\
2007 Jul  & $14.2^{+1.5}_{-1.9}$   & $0.173^{+0.049}_{-0.045}$   & $8.66^{+1.54}_{-1.94}$   & $0.120^{+0.067}_{-0.073}$   & $-0.023^{+0.013}_{-0.012}$  \\
2008 Jul  & $13.9^{+1.8}_{-2.6}$   & $0.046^{+0.023}_{-0.014}$   & $2.70^{+7.16}_{-1.07}$   & $0.074^{+0.024}_{-0.054}$   & $-0.016^{+0.009}_{-0.004}$  \\
2013 Aug  & $15.9^{+0.4}_{-0.6}$   & $0.052^{+0.005}_{-0.005}$   & $4.64^{+0.53}_{-0.66}$   & $0.012^{+0.009}_{-0.008}$   & $-0.002^{+0.001}_{-0.001}$  \\
2013 Sep  & $15.8^{+0.4}_{-4.6}$   & $0.028^{+0.004}_{-0.002}$   & $6.82^{+0.66}_{-0.87}$   & $-0.026^{+0.011}_{-0.008}$   & $0.001^{+0.001}_{-0.001}$  \\
2015 Sep  & $13.8^{+1.4}_{-1.0}$   & $0.206^{+0.040}_{-0.026}$   & $0.50^{+0.81}_{-0.38}$   & $-0.192^{+0.038}_{-0.068}$   & $0.028^{+0.010}_{-0.007}$  \\
\enddata
\end{deluxetable}

\section{Results}
\label{sec:results}

In this section we present the results of the \ion{Ca}{2} K flux analysis
and the search for periodicity in the residual fluxes for each epoch. 

\subsection{Mean \ion{Ca}{2} K fluxes and nightly variations}

The mean \ion{Ca}{2} K spectrum for each epoch is shown in the top row of
\autoref{fig:profs}. The 2015 July data shows the highest activity level, i.e.,
the largest core flux, and 2007 June is the lowest.   Also shown in
\autoref{fig:profs} are the mean absolute deviation (MAD) profiles for each
epoch (bottom row), which is a measure of how much the \ion{Ca}{2} K line core 
varies across the epoch. The MAD profile is constructed by taking the mean of the
absolute value of the nightly residual profiles. Note that epochs with the most
\ion{Ca}{2} K core emission do not necessarily show the strongest MAD profiles.

\begin{figure*}[htbp]
   \centering
   \includegraphics[scale=.75,clip,trim=9mm 55mm 10mm 25mm,angle=0]{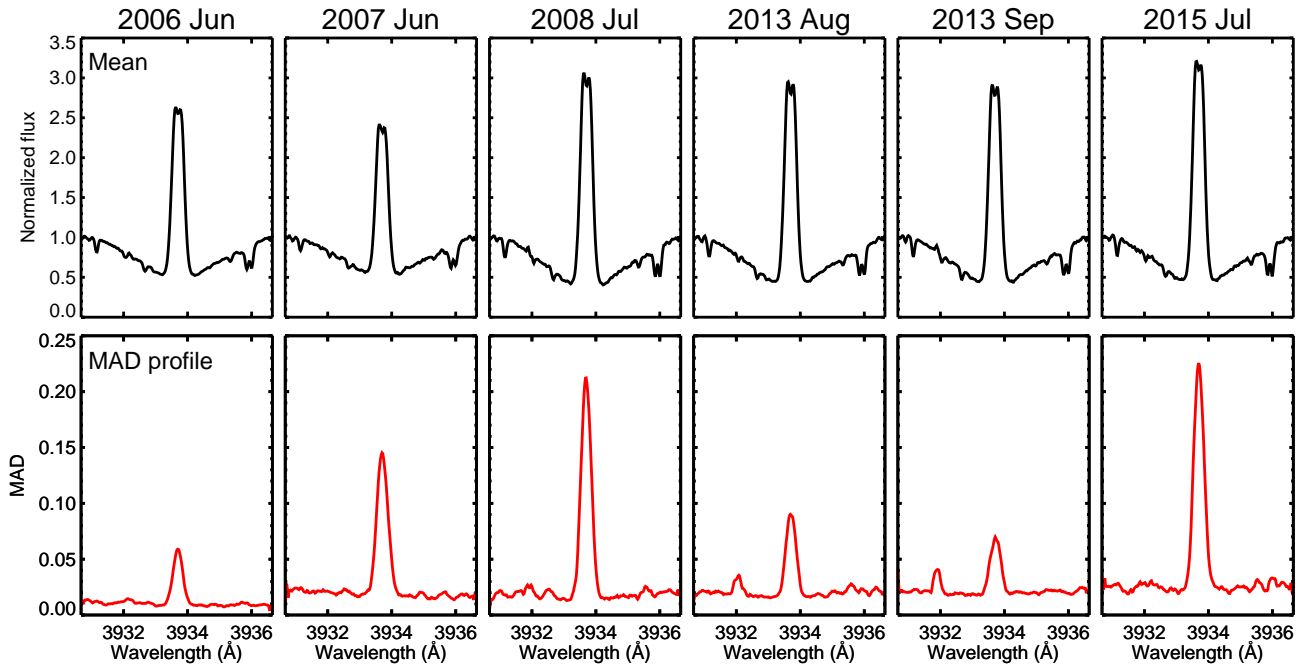}
   \figcaption{Master \ion{Ca}{2} K profiles for each epoch (top row) and
smoothed mean absolute deviation (MAD) profiles of the \ion{Ca}{2} K line core (bottom row).
The blips at $\approx 3932$ \AA\ in the 2008 July and 2013 August and September epochs
are the result of imperfect removal of a bad pixel in the NARVAL data. Epochs
with the largest \ion{Ca}{2} K emission do not necessarily show larger MAD profiles.
   \label{fig:profs}}
\end{figure*}

The spectra show a significant amount of nightly variability.
\autoref{fig:nvars} shows the integrated residual \ion{Ca}{2} K core flux for
all of the individual exposures. The intra-night variability is large on some
nights, spanning almost the full range of residual core flux for the entire
epoch, e.g., the fourth night (purple symbols) of 2006 June. The intra-night
variability also changes from night to night and epoch to epoch, highlighting
the dramatic nature of the constantly evolving active regions on the surface of
HD 189733.

We also examine how the intra-night variations relate to the overall
\ion{Ca}{2} K emission levels.  The residual core flux for each epoch is
calculated by integrating the spectra in the top row of \autoref{fig:profs}
across a 1 \AA\ band centered on \ion{Ca}{2} K. To quantify the magnitude of
the intra-night variability, we take the interquartile of the residual fluxes
in \autoref{fig:nvars}, hereafter referred to as IQR, for a single night. We
then average these nightly IQR values to find the mean IQR for the epoch. The
IQR uncertainties are the standard deviation of the mean for that epoch. The
mean IQR values are shown as a function of the \ion{Ca}{2} K core flux in the
left panel of \autoref{fig:vars}. We find no trend for the IQR as a function of
\ion{Ca}{2} K core flux.

\begin{figure*}[htbp]
   \centering
   \includegraphics[scale=.83,clip,trim=5mm 75mm 0mm 60mm,angle=0]{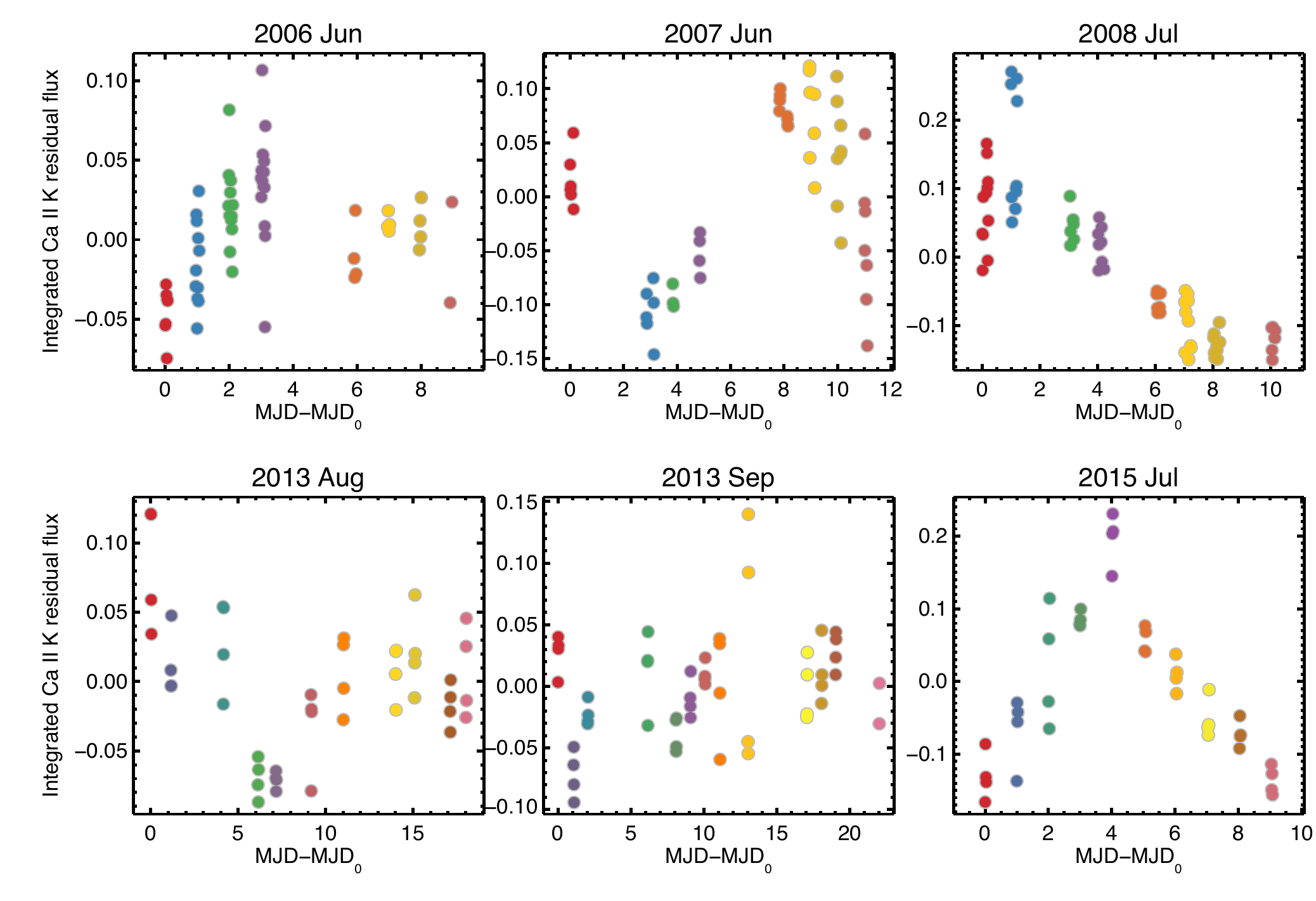}
   \figcaption{Integrated residual \ion{Ca}{2} K core flux for individual
spectra as a function of time. Spectra from the same night are plotted
with the same color. Note the different y-axis scaling for each epoch.
The spectra show large nightly variations in
the \ion{Ca}{2} K core flux and the magnitude of the variations changes
night to night and epoch to epoch.
   \label{fig:nvars}}
\end{figure*}

There is, however, a correlation between mean IQR and the integrated MAD flux,
which is shown in the right panel of \autoref{fig:vars}: epochs with larger MAD
fluxes tend to exhibit larger IQR values. In other words, epochs with greater
average \ion{Ca}{2} K core flux variations across the entire epoch (e.g., from
rotational and any orbital modulation) are the same epochs which show the
largest average intra-night variability. This suggests that average
night-to-night variations are a better proxy for how active the star is on
hours-long timescales than the absolute core flux level of the epoch's mean
\ion{Ca}{2} K spectrum.

\begin{figure*}[htbp]
   \centering
   \includegraphics[scale=.72,clip,trim=10mm 40mm 17mm 65mm,angle=0]{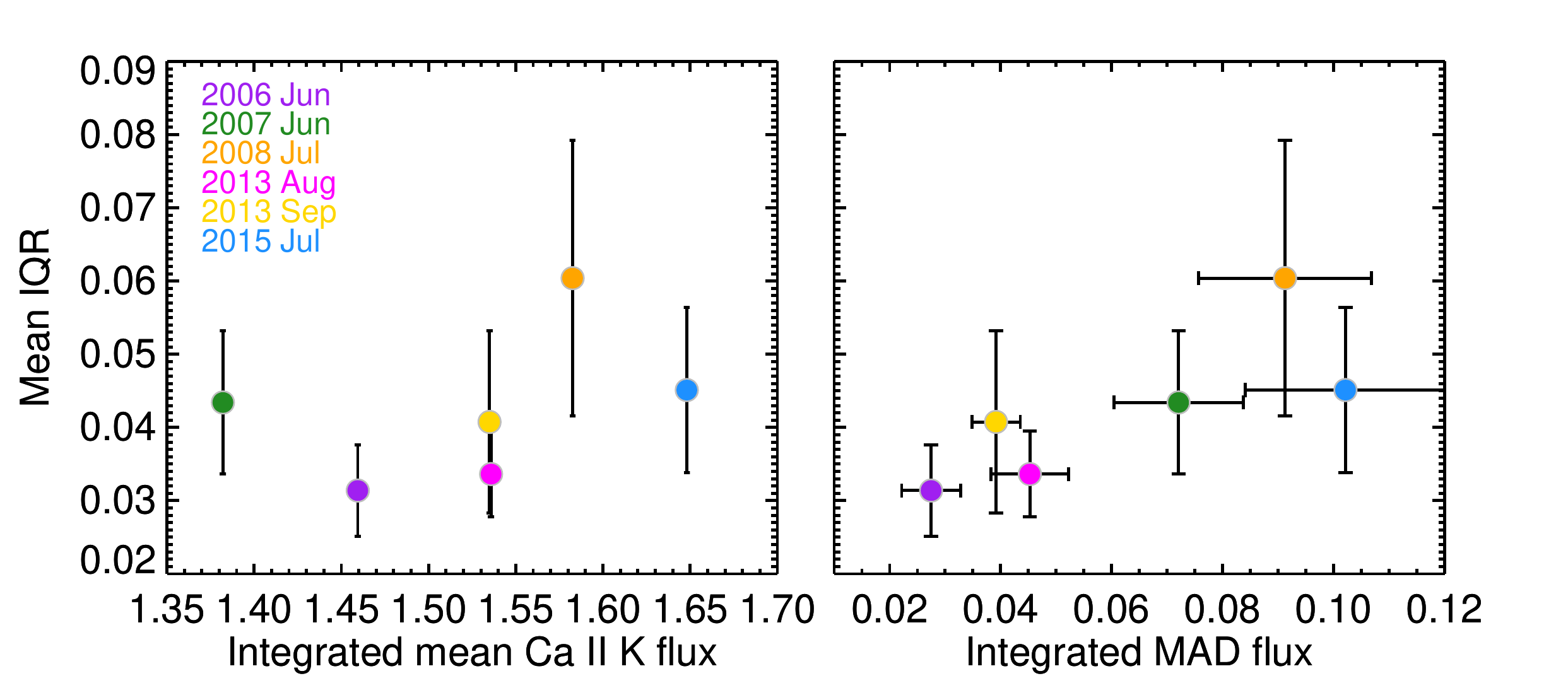}
   \figcaption{Mean nightly interquartile residual fluxes (IQR) versus the
integrated \ion{Ca}{2} K core flux (left panel) and integrated MAD flux (right panel)
for each epoch. The uncertainties for the
\ion{Ca}{2} K core fluxes are smaller than the plot symbols. There is no noticeable trend
with \ion{Ca}{2} K core flux, suggesting that HD 189733 exhibits stochastic
nightly variability at a similar magnitude regardless of the absolute activity level.
However, a trend can be seen in the right panel: epochs with overall larger
changes in \ion{Ca}{2} K core flux, i.e., larger MAD fluxes, tend to show
greater intra-night variability, or IQR values.
   \label{fig:vars}}
\end{figure*}

\subsection{Periodic variations in the \ion{Ca}{2} K residuals}
\label{sec:psearch}

The \ion{Ca}{2} K residual profiles and flux as a function of time are shown in
\autoref{fig:phase}. The rotation fits from \autoref{sec:rotvar} are over
plotted in the middle rows and the residual flux minus the rotation fits are
given in the bottom rows.  Uncertainties for the residual flux are derived by
summing in quadrature the normalized flux uncertainties of the residual
spectrum. Note that the normalized flux uncertainty is estimated from the
spectrum adjacent to the line core, where the signal-to-noise is $\approx
2\times$ lower than in the line core HD 189733 (see \autoref{fig:profs}). Thus
the residual flux uncertainties are conservative. 

\begin{figure*}[htbp]
   \centering
   \includegraphics[scale=.85,clip,trim=13mm 25mm 10mm 5mm,angle=0]{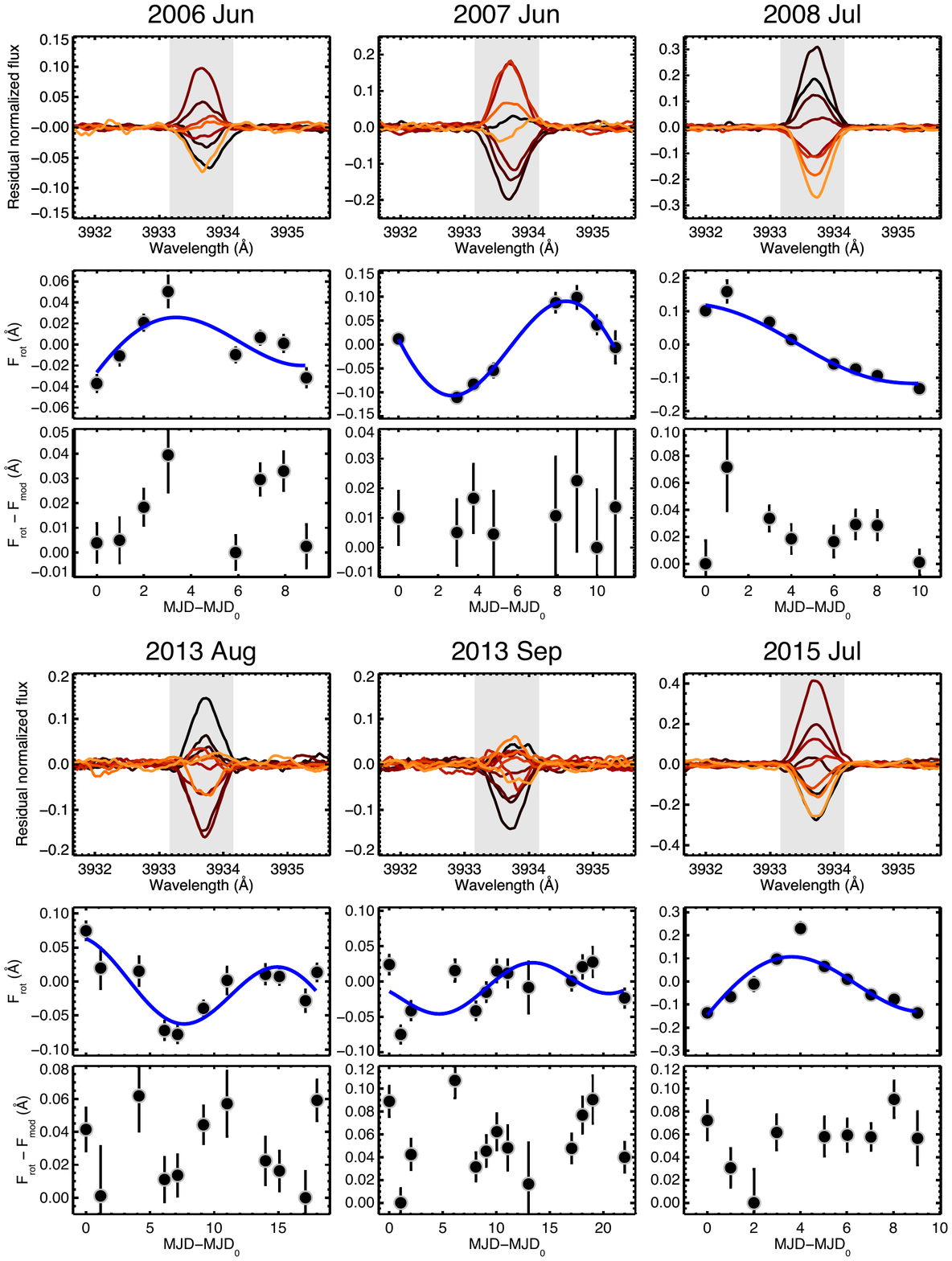}

   \figcaption{Residual \ion{Ca}{2} K profiles (top rows), raw residual flux
curves (middle rows), and residual flux with rotational modulations removed
($F_\text{rot} - F_\text{mod}$).  The residual line profiles are smoothed
for clarity. The gray shaded regions indicate the residual
flux integration range. Spectrum colors move from black for earlier dates to
yellow-orange for later dates. The blue lines are the rotational fits to the
raw flux curves. Note that the y-axis scaling is different for each epoch and
that the $F_\text{rot} - F_\text{mod}$ values have been shifted so that the
lowest value is zero.  \label{fig:phase}}

\end{figure*}

We employed the Lafler-Kinman statistic to search for periodicity in
the rotation-subtracted \ion{Ca}{2} K residuals in \autoref{fig:phase}. The
Lafler-Kinman (LK) statistic is a non-parametric string-length method for
evaluating periodicity within time series data \citep{clarke02}.
\citet{clarke02} explored the behavior of the LK statistic, which we refer to
as \tp\ \citep[see equation 3 of][]{clarke02}, as a function of the
number of time series points $N$ and the signal-to-noise ratio (SNR) of the
data. They derived values of \tp\ which describe the maximum value for
which a test period can be considered significant at some confidence level.
These maximum \tp\ values are a function of $N$ and the SNR. Here we
interpolated among the 90\%, 95\%, and 99\% \tp\ values from Table 1
of \citet{clarke02} for the appropriate values of $N$ and SNR of our data. Note
that for small $N$ there is sometimes no value of \tp\ corresponding
to some confidence level. 

We performed the period search within the range 1 day $\leq P \leq $
max(T-T$_0$) where T-T$_0$ is the number of nights spanning the entire epoch.
We initially chose a period resolution of 0.01 days. Calculating \tp\ requires
ordering the data by increasing phase. We use the inferior conjunction (i.e.,
mid-transit) BJD $T_0 = 2453955.525511$ \citep{baluev15} for the phase
calculations. The \tp\ values for each epoch are shown in
\autoref{fig:psearch}. The red, orange, and purple lines mark the 90\%, 95\%,
and 99\% confidence values of \tp\ for each case. Note that the small
number of points and poor SNR for 2007 June ($N = 8$, $\text{SNR} \approx 0.7$)
makes it impossible to assign a confidence of $\geq 90\%$ to \textit{any} value
of \tp; thus no significant periods are reported for this epoch. The
2013 September is the only data set for which a 99\% confidence value of
\tp\ can be determined.

It is important to reiterate that the confidence levels do not imply
that some period with a low \tp\ value is a physical period; they instead give
the probability with which that period is detected in the data set. Thus it is
possible to have spurious periods with \tp\ values below some high confidence
level. Evaluation of the likelihood that a similarly significant period would
be detected by chance must be performed separately, which we do in the next
section.   

\begin{figure*}[ht!]
   \centering
   \includegraphics[scale=.75,clip,trim=13mm 10mm 10mm 15mm,angle=0]{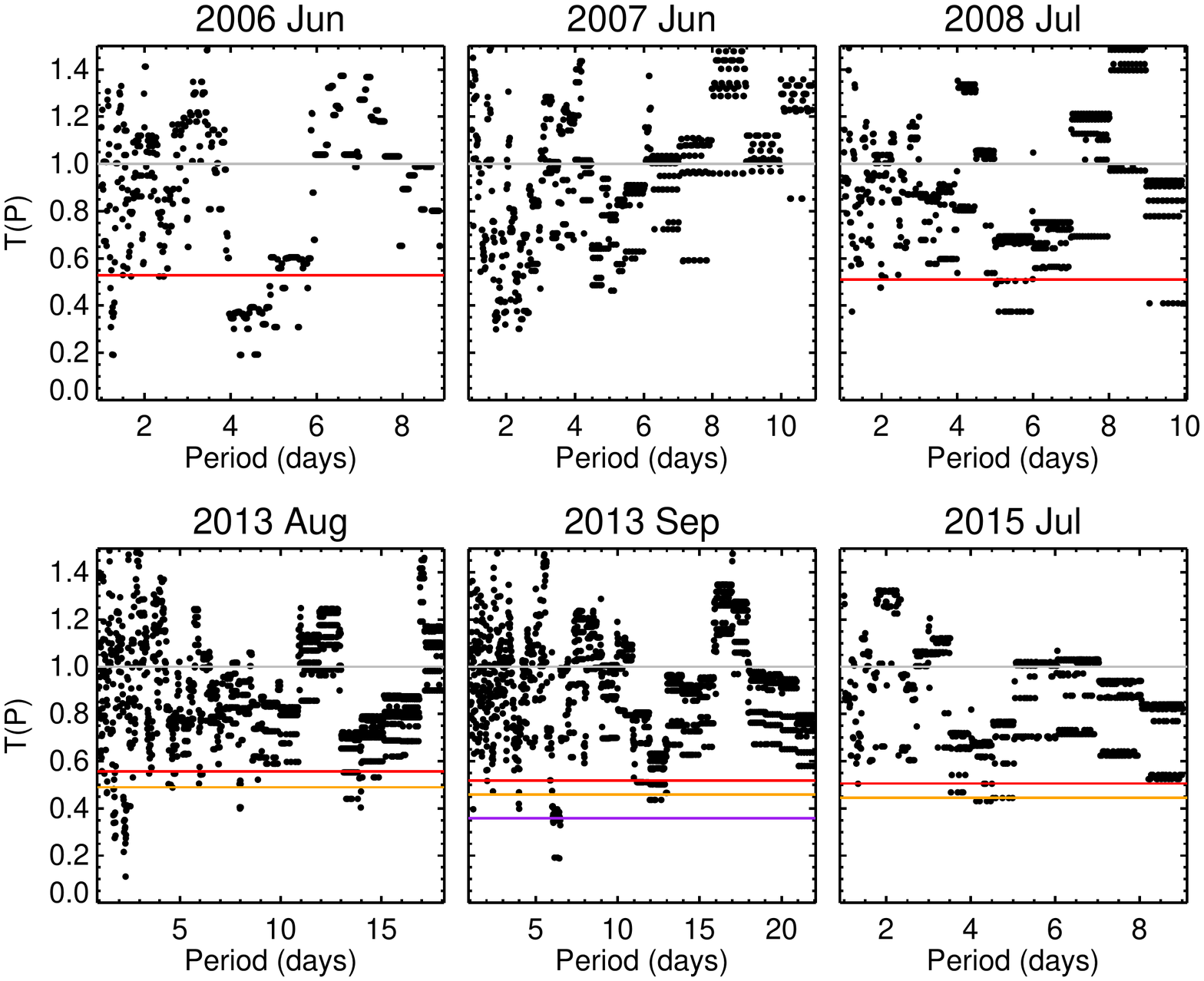}

   \figcaption{$T(\text{P})$ as a function of period for each epoch. The
$T(\text{P})$ values indicating a period at 90\%, 95\%, and 99\% confidence are
marked with the horizontal red, orange, and purple lines, respectively. No
periods with confidence greater than 90\% can be determined for the 2006 June
and 2008 July epochs, while no periods with significance at $> 90\%$ can
be determined at all for the 2007 June epoch. \label{fig:psearch}}

\end{figure*} 

\begin{figure*}[htbp]
   \centering
   \includegraphics[scale=.75,clip,trim=13mm 10mm 10mm 30mm,angle=0]{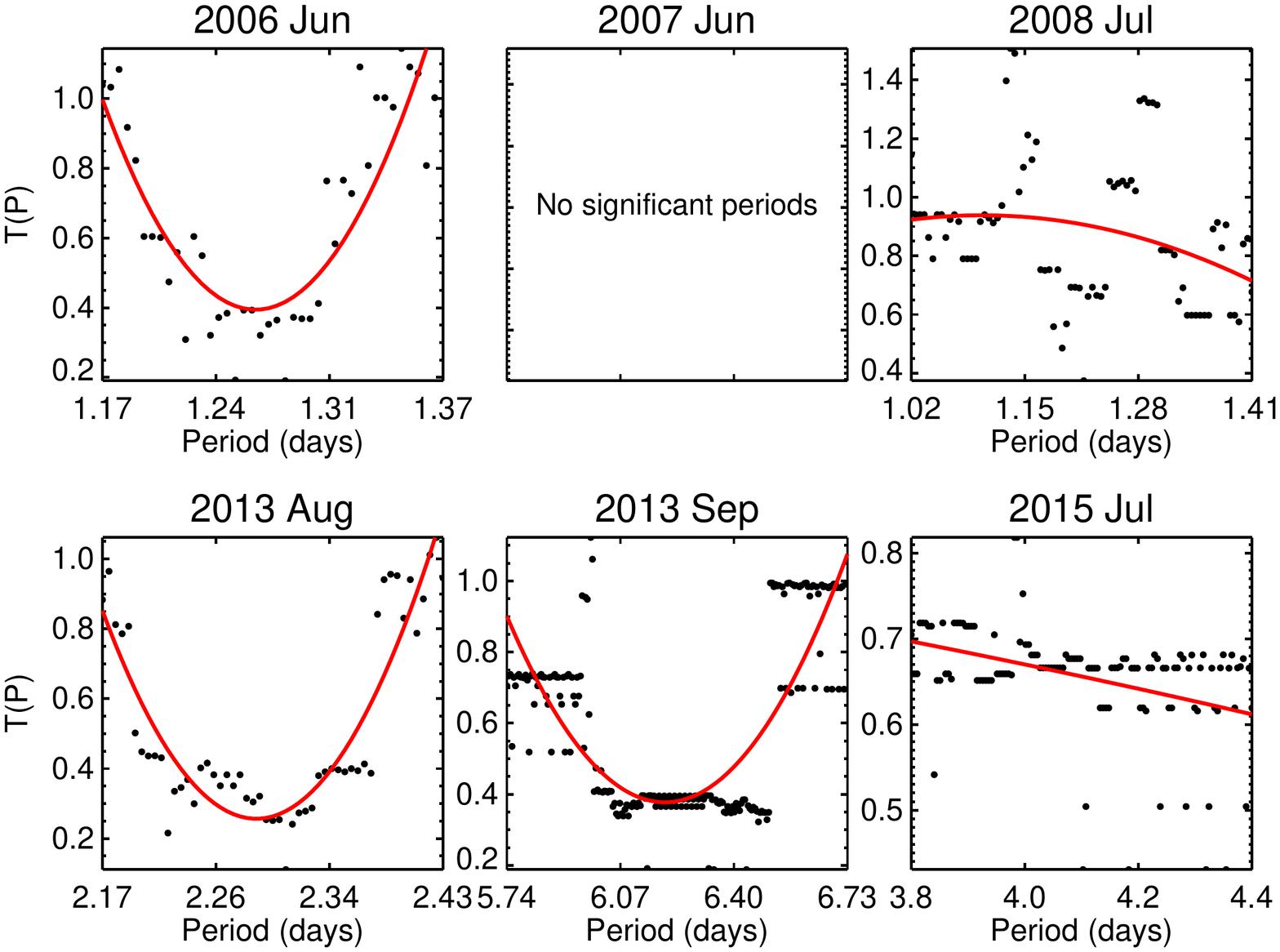}

   \figcaption{$T(\text{P})$ values zoomed in on the lowest $T(\text{P})$
values from \autoref{fig:psearch}. The period space around the minima are
explored with a resolution of 0.005 days. The best-fit parabola to the \tp\
minima are over-plotted in red. Note that the y-axis scale is different for
each epoch.  Only the epochs of 2006 June and 2013 August show well-defined
minima; 2013 September has a much broader minimum and is weakly described by a
parabolic fit. The best-fit period for 2013 August is consistent with the
planet's orbital period.  \label{fig:pzoom}}

\end{figure*}

\subsection{Statistical tests of detected periods}
\label{sec:stattests}

In order to determine the uncertainties of the lowest \tp\ periods in
\autoref{fig:psearch}, we adopted the procedure from \citet{fernie89}
\citep[see also][]{kwee56}. The procedure involves finely sampling the
period space near the \tp\ minimum and fitting a parabola to the \tp\ values in
the vicinity. The quadratic coefficients are then used to calculate the period
and its uncertainty.  The zoomed-in period-\tp\ space is shown in
\autoref{fig:pzoom} and the best-fit periods and associated uncertainties are
given in \autoref{tab:periods}. The 2008 July and 2015 July epochs show no
smooth minima and thus no reliable fits. We reject these low \tp\ values as
spurious. The 2006 June, 2013 August, and 2013 September data display
reasonable minima, although the 2013 September space is fairly broad. The
lowest \tp\ value found for any epoch is \tp $= 0.11$ for 2013 August. The
best-fit period of $2.29\pm0.04$ days for 2013 August is consistent at $\approx
2\sigma$ with the planet's orbital period of $P_\text{orb} \approx 2.219$ days.

\begin{deluxetable}{lccc}
\tablewidth{1.99\textwidth}
\tablecaption{Lowest T(P) periods and their uncertainties\label{tab:periods}}
\tablehead{\colhead{}&\colhead{}&\colhead{Period}&\colhead{}\\
\colhead{Observation epoch}&\colhead{min($T(\text{P})$)}&\colhead{(days)}&\colhead{Detection significance}}
\colnumbers
\tabletypesize{\scriptsize}
\startdata
 2006 Jun & 0.19 & $1.26 \pm 0.04$ & $>90\%$ \\
 2007 Jun & \nodata & \nodata & \nodata \\
 2008 Jul & 0.37 & \nodata & \nodata \\
 2013 Aug & 0.11 & $2.29 \pm 0.04$ & $>95\%$ \\
 2013 Sep & 0.19 & $6.20 \pm .17$ & $>99\%$ \\
 2015 Jul & 0.43 & \nodata & \nodata \\
\enddata
\end{deluxetable}

\autoref{fig:phres} shows the \ion{Ca}{2} K residuals phased to the best-fit
periods from \autoref{tab:periods} for 2006 June, 2013 August, and 2013
September. The 2006 June and 2013 August residuals have very similar forms,
e.g., truncated sine curves, with peak fluxes near $\phi \approx 0.7-0.8$ for
2006 June and $\phi \approx 0.0-0.1$ for 2013 August. The 2013 September
residuals show a gradual increase in flux and then a sharp drop near $\phi
\approx 0.5$. 

\begin{figure*}[htbp]
   \centering
   \includegraphics[scale=.71,clip,trim=5mm 25mm 20mm 105mm,angle=0]{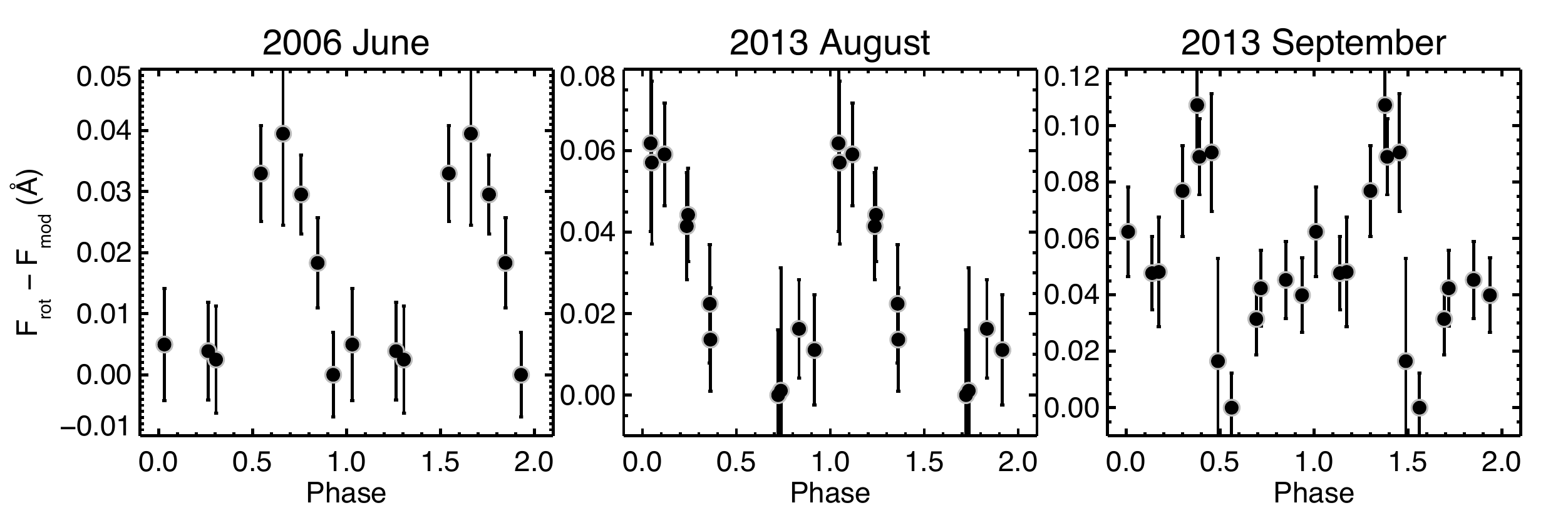}

   \figcaption{Residual \ion{Ca}{2} K fluxes, with rotational modulations
removed, phased to the best-fit periods from \autoref{tab:periods}. The data
is repeated in phase for display purposes. The 2006 June and 2013 August residuals
show peak fluxes at approximately the same phase. \label{fig:phres}}

\end{figure*}

In order to test the significance of the best-fit periods, we performed
statistical assessments of how likely it is to obtain a set of residual fluxes
with a similar or better value of \tp. In other words, how likely is it to
measure $\tp \leq \text{min}(\tp)$ from a randomly generated set of residual
fluxes? To do this we looked at three cases: 1. normally distributed fluxes
with standard deviation and mean equal to the measured fluxes; 2.  uniformly
distributed fluxes bounded by the minimum and maximum of the measured values;
and 3. randomly re-ordered values of the measured fluxes. The observation dates
are kept constant. For each case we generated $N=5000$ random flux vectors and
performed the same \tp\ analysis for each vector as was done for the observed
data. We then took the minimum \tp\ value from the period analysis and compared
all of these \tp\ values to the observed minimum \tp\ value. The $p$-value for
each case is then the fraction of repetitions which produce $\text{min}(\tp)
\leq \text{min}(\tp)_\text{obs}$.

The \tp\ histograms and derived $p$-values for the epochs with
significant periods are shown in \autoref{fig:chi} and the $p$-values are
summarized in \autoref{tab:pvals}. The 2006 June epoch has fairly large
$p$-values for all of the test methods, suggesting that the detected period is
not robust.  This is mainly the result of the small number of observations
($N=8$). Although not as high, the 2013 September epoch did not produce a
$p$-value less than 1\% and did particularly poorly in the Uniform test case,
with $p = 0.052$. The most robust signal by far is from the 2013 August data:
$p<0.01$ in all of the test cases, providing evidence that $<1\%$ of random
residual flux measurements will result in a value of \tp\ as low as that
measured from the data. We thus consider the 2013 August signal to be the only
significant detection and do not further discuss the marginal cases of 2006
June and 2013 September.   
  
\begin{figure*}[htbp]
   \centering
   \includegraphics[scale=.85,clip,trim=33mm 45mm 10mm 15mm,angle=0]{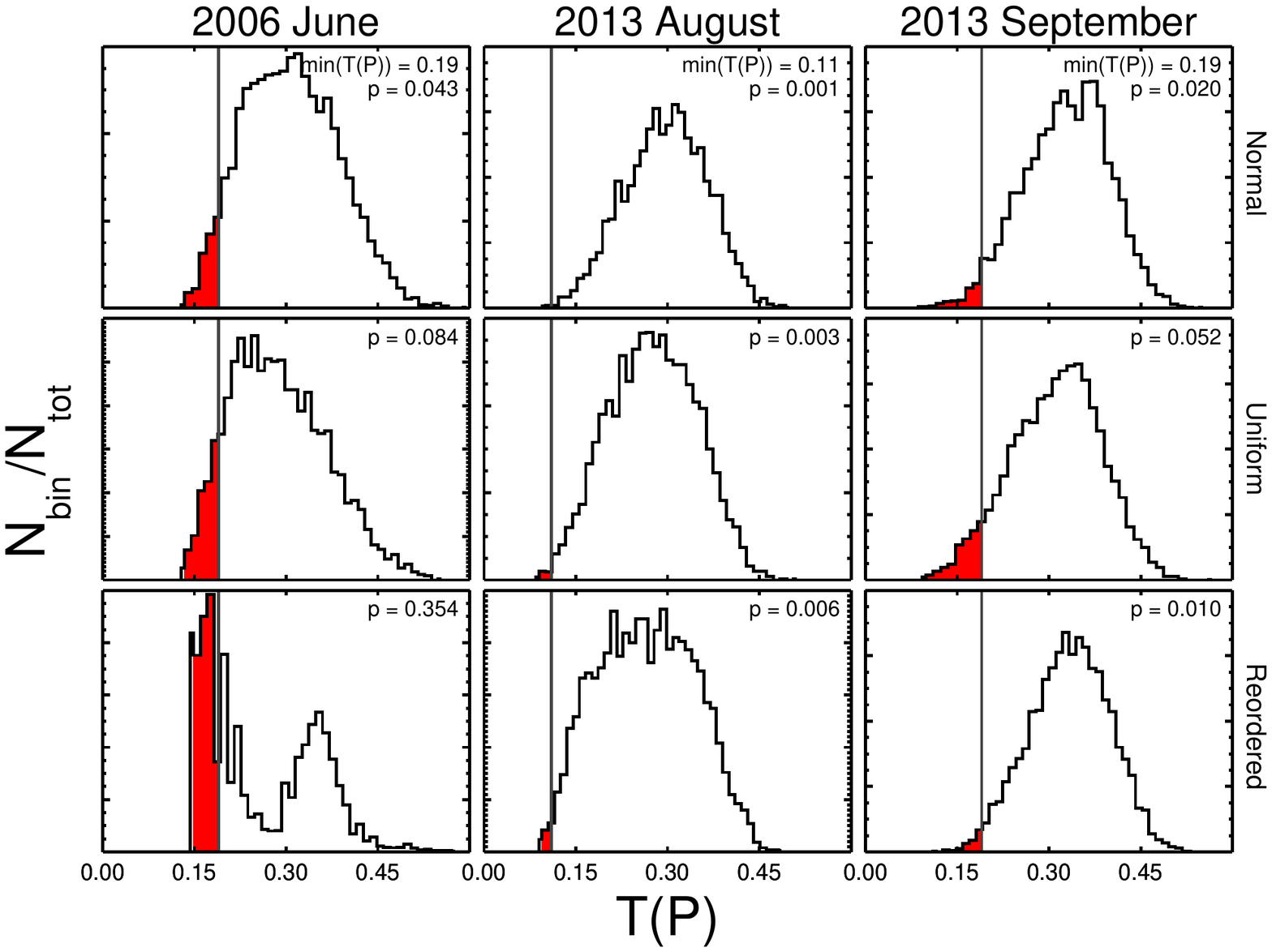}

   \figcaption{\tp\ histograms for random flux tests. The rows represent the
different methods: random draws from a normal distribution (first row), random
draws from a uniform distribution (second row), and randomly reordered flux
vectors (third row). Red filled regions show $\tp \leq
\text{min}(\tp)_\text{obs}$.  The $\text{min}(\tp)_\text{obs}$ value is marked
with a vertical gray line and labeled in the upper-right of the top panels.
The $p$-values are also given for each test. The only epoch with a consistently
low $p$-value is 2013 August, suggesting a robust period detection. \label{fig:chi}}

\end{figure*}

\begin{deluxetable}{lccccc}
\tablewidth{1.99\textwidth}
\tablecaption{Minimum \tp\ and statistical test $p$-values\label{tab:pvals}}
\tablehead{\colhead{}&\colhead{}&\multicolumn{3}{c}{$p$-values}&\colhead{}\\
\colhead{Observation epochs}&\colhead{min(\tp)}&\colhead{Normal}&\colhead{Uniform}&
\colhead{Reordered}&\colhead{Mean}}
\colnumbers
\tabletypesize{\scriptsize}
\startdata
 2006 Jun & 0.19 & 0.043 & 0.084 & 0.354 & 0.160\\
 2013 Aug & 0.11 & 0.001 & 0.003 & 0.006 & 0.003\\
 2013 Sep & 0.19 & 0.020 & 0.052 & 0.010 & 0.027\\
\enddata
\end{deluxetable}

\section{Discussion}
\label{sec:discussion}

We analyzed periodicity in the rotation-subtracted \ion{Ca}{2} K residual flux
for six observations epochs of HD 189733. We found one statistically
robust period: $P = 2.29\pm0.04$ days for the 2013 August epoch.
This period is consistent with the orbital period of HD 189733 b.

Magnetic SPI signals are predicted to be modulated at the planet's
orbital or beat period, where the planet continuously excites Alfv\'en waves
\citep{preusse06,kopp11,strugarek16} or experiences reconnection events with
the stellar magnetic field as it moves through the magnetosphere
\citep{lanza11,lanza12}. Thus the detection of a robust period very near the
planet's orbital period is an indication that we are in fact measuring the
effects of magnetic SPI in the system.  

Besides the evidence of the SPI signal itself, the 2013 August variations are
interesting for a variety of reasons. The second lowest \tp\ value in
\autoref{fig:pzoom} is at 2.22 days, almost exactly the orbital period of HD
189733 b \citep[$P_\text{orb} = 2.218575$ days; ][]{baluev15}. The
rotationally-subtracted \ion{Ca}{2} K residuals phased to the planet's orbital
period are shown in \autoref{fig:orbphase}. If we assume that the signal
is strongest when the projected area of the emitting region is largest, i.e.,
when the SPI hot spot crosses the line-of-sight, this corresponds to a phase
lead of $\Delta\phi \approx 40^\circ$ ahead of the planet. This is roughly
consistent with the phase lead of $\Delta\phi = 53^\circ$ for HD 189733 found by
\citet{lanza12} for a model assuming a non-linear force-free stellar magnetic
field.

\begin{figure}[htbp]
   \centering
   \includegraphics[scale=.51,clip,trim=45mm 35mm 30mm 45mm,angle=0]{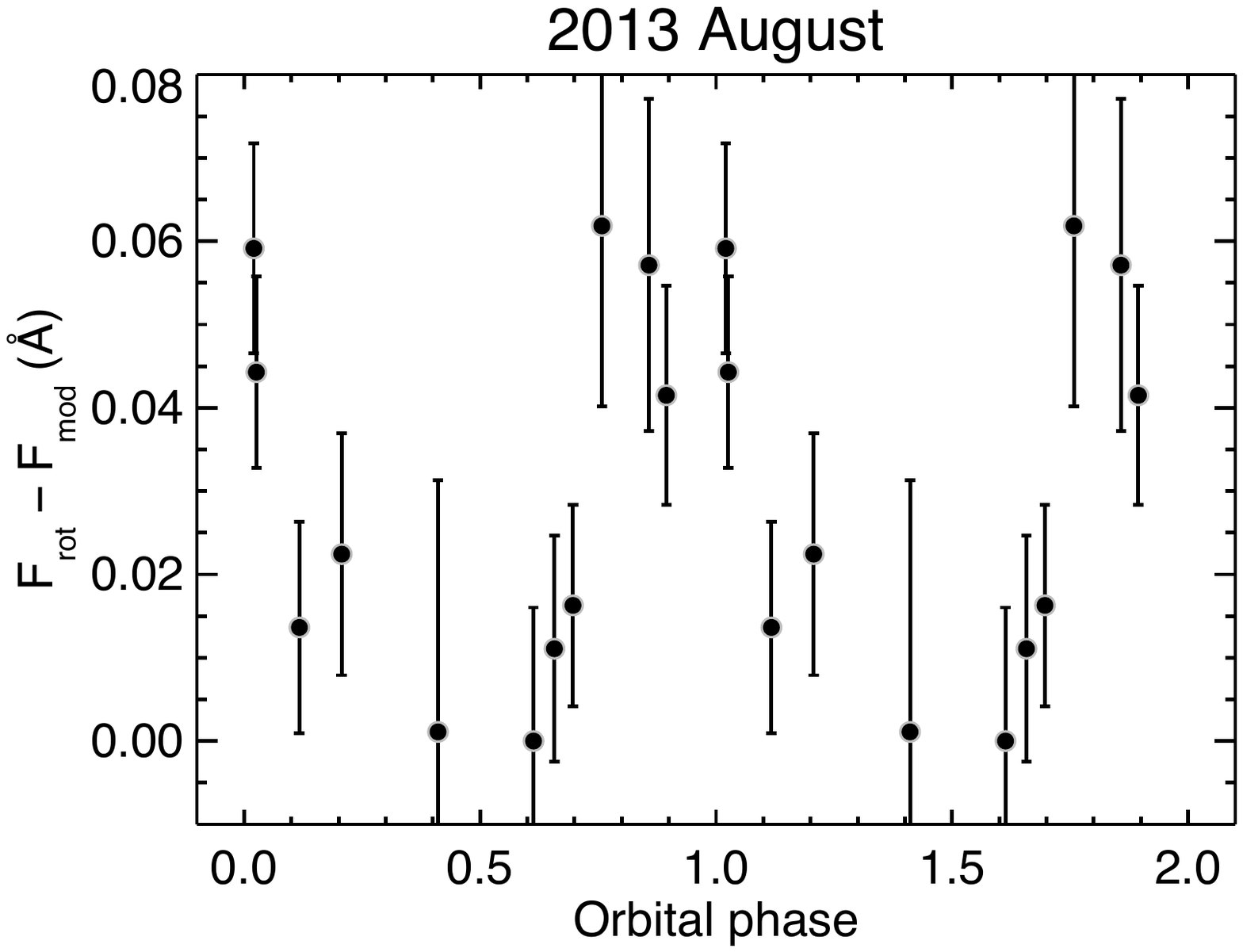}

   \figcaption{Rotationally-subtracted \ion{Ca}{2} K residual fluxes as
a function of the precise orbital period of HD 189733 b. The average phase
of the maximum fluxes is $\approx 0.9$, which translates to a phase lead 
by the SPI hotspot of $\Delta\phi \approx 40^\circ$. \label{fig:orbphase}}

\end{figure} 

\citet{fares17} calculated the stellar magnetic field strength and
configuration of HD 189733 for all of the epochs examined here, with the
exception of the 2006 June epoch, which was presented by \citet{moutou07}. They
find that the stellar field changes significantly, even on timescales of a few
rotation periods. They extrapolate the stellar field strength to the planet's
orbital distance and find average field strengths of $B^\text{orb}_* = 16, 23,
39, 31$, and 18 mG for the epochs of 2007 June, 2008 July, 2013 August, 2013
September, and 2015 September, respectively \citep[see Table 4 of ][]{fares17}.
We note that although no field extrapolation exists for the 2006 June epoch,
the surface field strength is low ($B_* = 18$ G), ruling out a large field
strength at the planet's orbit. \citet{fares17} also show that the planet can
move through drastically different coronal conditions over a single orbit
\citep[see also ][]{llama13}.  

The $B^\text{orb}_*$ value for 2013 August may offer a clue as to why this
epoch shows evidence for magnetic SPI: the power in magnetic SPI scales as $P
\propto v_\text{rel} R_\text{p}^2 B_\text{p}^{2/3} B_*^{4/3}$
\citep[e.g.,][]{lanza09,lanza12} where $v_\text{rel}$ is the relative velocity
between the planet and stellar magnetic field lines, $R_\text{p}$ is the
planet's radius and $B_\text{p}$ is the planetary magnetic field, suggesting
that a larger stellar magnetic field strength at the planet's orbit should
produce stronger SPI emission. Assuming the other system parameters are
constant, the value of $B^\text{orb}_*$ for 2013 August implies that the
available power for magnetic SPI is $\approx 2.3\times$ greater on average than
for the other epochs. Thus the larger magnetic field strength may be an
important contributor to the observed SPI signature. 

\section{Summary and Conclusions}
\label{sec:conclusion}

We analyzed six archival high resolution \ion{Ca}{2} K data sets (58 total
nights) for the hot Jupiter system HD 189733 in order to search for signatures
of magnetic star-planet interactions. After removing the rotational modulations
we find statistically significant ($\bar{p} = 0.003$; $>99\%$) variations for
the 2013 August data set at a period of $2.29\pm0.04$ days, consistent with the
orbital period of HD 189733 b; none of the other epochs show variations with a
significance greater than $\approx 95\%$, which is not unexpected given the
high activity level of HD 189733 and the short timescales over which the
stellar magnetic field changes strength and morphology \citep{fares17}. 

The SPI signal reported here has characteristics that are consistent with the
non-linear force-free magnetic SPI model suggested by \citet{lanza12},
specifically the phase offset of $\Delta\phi \approx 40^\circ$ between the
planet and magnetic interaction. The absence of SPI signals in the majority of
the analyzed epochs illustrates the fickle nature of these signals
\citep[e.g.,][]{shkolnik08}, which depend on the magnetic field strength
and configuration of the host star at the time of observations and are likely
obscured, even when present, to some extent by stochastic stellar variability.
Our results reinforce the need for rigorous monitoring of any potential SPI
systems over many planetary orbits in order to maximize the opportunity of
catching the star and planet in a favorable configuration. 

\bigskip

{\bf Acknowledgments:} We thank the referee for their consideration of the
manuscript, which helped improve the analysis and results. We thank Gordon
Walker and David Bohlender for their valuable input.  This work is supported by
NASA Origins of the Solar System grant No. NNX13AH79G (PI: E.L.S.). This work
has in part been carried out in the frame of the National Centre for Competence
in Research PlanetS supported by the Swiss National Science Foundation (SNSF).
V.B. acknowledges the financial support of the SNSF. This project has received
funding from the European Research Council (ERC) under the European Union's
Horizon 2020 research and innovation programme (project {\sc Four Aces}; grant
agreement No 724427).This work has made use of NASA's Astrophysics Data System.

\end{document}